\documentclass[12pt]{article}
\title{Point Form Electrodynamics and the Construction of Conserved
Current Operators}
\author{W. H. Klink\\Department of Physics and Astronomy\\ University of
Iowa, Iowa City, Iowa, USA}
\begin{document}
\maketitle
\begin{abstract}
A general procedure for constructing conserved electromagnetic current
operators is given. The four-momentum operator in point
form relativistic quantum mechanics is written as the sum of
hadronic, photon, and electromagnetic four-momentum operators, where the
electromagnetic four-momentum operator is generated from a vertex operator,
in which a conserved current operator is contracted with the
four-vector potential operator.  The current operator is the
sum of free, dynamically determined and model dependent operators, the
sum of the first two constructed so as to be
conserved with respect to the hadronic interactions. It is shown that the 
dynamically determined operator carries charge.
The model dependent operator is a many-body current operator,
formed as the commutator of an antisymmetric operator with
the hadronic four-momentum operator.  It is shown that such
an operator is conserved and does not renormalize the
charge.\footnote{PACS, 11.40.Dw, 13.40.Gp. }.
\end{abstract}
\section{Introduction}
A constantly recurring issue in few-body quantum mechanics
concerns the construction of conserved electromagnetic current operators. 
One-body current operators, such as those constructed in the second paper
in this series \cite{a}, are conserved when there are no hadronic
interactions.  But in the presence of hadronic interactions they are no
longer conserved.

 Such issues already arise in nonrelativistic quantum mechanics. 
Let
$H$ be the hadronic Hamiltonian, $\rho({\vec x})$ the charge density
operator, and ${\vec J}({\vec x})$ the current density operator.  The
electromagnetic charge and current operators in the Heisenberg
representation are then given by
\begin{eqnarray}
\rho(t,{\vec x})&=&{e^{iHt}} {\rho({\vec x})}{e^{-iHt}}\nonumber\\
{\vec J}(t,{\vec x})&=&{e^{iHt}}{{\vec J}({\vec x})}{e^{-iHt}},\nonumber
\end{eqnarray}
and the continuity equation, $\partial\rho(t,{\vec x})/\partial t
+\nabla\cdot {\vec J}(t,{\vec x})=0$ can be written as the commutator
equation,
\begin{eqnarray}
i[H,\rho({\vec x})]+\nabla\cdot {\vec J}({\vec x})=0.\nonumber
\end{eqnarray}
If the charge density operator $\rho({\vec x})$ is chosen to be a one-body
operator and the interaction terms in $H$ do not commute with the charge
density operator,  the current density operator will contain not only
one-body terms, but two (or more) body terms in order that the
continuity equation be satisfied.  But this does not fix the current
density operator, since the continuity equation only involves the
divergence of the current density operator.  If the current density
operator is decomposed into so-called parallel and perpendicular
components, then only the parallel component is fixed by the continuity
equation, and the perpendicular component is left unspecified,
resulting in what are called model dependent currents \cite{b}.

It will be shown that the same situation holds in relativistic quantum
mechanics, with the additional stipulation that, in the presence of hadronic
interactions, there must always be two (or more)-body currents, in order to
be relativistically covariant.  Moreover, the decomposition of the current
operator into two parts, one of which is divergenceless, is not 
relativistically invariant.  In the point form of relativistic quantum
mechanics, the form used in this series of papers, one-body current
operators with the correct Lorentz and other transformation properties were
given in the previous paper for arbitrary spin particles \cite{a}; but, in
the presence of hadronic interactions, in order to satisfy current
conservation, there must nonetheless be two or more body operators.  This is
most readily seen in the Breit frame, where the zero component of the four
vector current operator gives the electric form factors, and the one and
two components give the magnetic form factors.  The third component in the
Breit frame should be zero in order for current conservation to hold.  But
a direct calculation shows that this is not the case if there are only
one-body currents \cite{c}.  Only if (implicit) two or more-body currents
are included, will current conservation hold.

In quantum field theory it might seem that once the Lagrangian for the
fields is given, the electromagnetic current operator is uniquely
given.  But even here there are problems having to do with satisfying the
field equations.  Let $\mathcal{L}$ be the Lagrangian consisting of
polynomials in free field operators, written as
$\mathcal{L}=\mathcal{L}_{matter}+\mathcal{L}_{\gamma}+\mathcal{L}_{em}$,
where
$\mathcal{L}_{matter}$ is the matter Lagrangian, for example for pions and
nucleons, $\mathcal{L}_{\gamma}$ is the photon Lagrangian, and
$\mathcal{L}_{em}$ is the electromagnetic Lagrangian, in which the charged
matter fields are coupled to the photon field.  Usually the electromagnetic
part of the Lagrangian is given through minimal substitution.  Then
Noether's theorem says that if
$\mathcal{L}$ is invariant under some internal symmetry transformation, for
example a constant phase transformation, there exists a quantity
$J_{\mu}(x)$ which is conserved if the field equations are satisfied. 
But solving the field equations in the presence of interactions is very
difficult, even for the electron-photon system, let alone for the
pion-nucleon system to which the photon field must be coupled \cite{d}.

  One of the goals of this paper is to show how to construct  conserved
electromagnetic current operators in the presence of hadronic interactions.
The current operators will be the sum of one-body and many-body current
operators, pieces of which will be the analogues of the model dependent and
model independent non-relativistic currents. The one-body current operators
have already been constructed in the second paper in this series;  however,
since they are not conserved with respect to the hadronic interactions,
they will be modified by adding pieces that make them conserved with
respect to the hadronic interactions.  Such a modification is analogous to
the model independent addition to the one-body currents in the
non-relativistic case.  

The construction of the analog of the model dependent currents in
non-relativistic quantum mechanics arises from the translational
covariance properties of current operators.  Many-body currents will be
constructed that are conserved and carry zero charge, by analyzing, in
section 3, the structure of point form electrodynamics.  Such currents are
given as the commutator of a field tensor with the hadronic four-momentum
operator;  the model dependence arises because the field tensor can be
chosen from a wide variety of operators, including one-body current
operators, as well as Feynman diagrams with external photon lines.

In section 2 the main features of point form relativistic quantum
mechanics are quickly reviewed, while in section 3 the ideas of point form
electrodynamics are laid out.  In particular it is shown that matrix
elements of the current and electromagnetic field operators recover the
classical Maxwell equations.  Then in section 4 the analog of the model
independent and in section 5 the analog of model dependent currents are
constructed.  The sum of free, model independent and model dependent
currents
 gives the total current operator, which, when coupled to the the
photon field, generates the electromagnetic vertex.
\section{Point Form Relativistic Quantum Mechanics}
In point form relativistic quantum mechanics all interaction are put in
the four-momentum operator.  The Lorentz generators contain no
interactions so that the point form is manifestly Lorentz covariant.  In
fact the Lorentz generators will almost never be used; instead
global Lorentz transformations are used to define the transformation
properties of operators and states.  In particular the Lorentz
transformation properties of photons, worked out in reference \cite{e},
are needed for the electromagnetic vertex.  If
$U_\Lambda$ is a unitary operator representing the Lorentz
transformation $\Lambda$, then the four-momentum operator must satisfy the
following point form equations:
\begin{eqnarray}
[P_\mu,P_\nu]&=&0\\
U_\Lambda P_\mu U_\Lambda ^{-1}&=&(\Lambda^{-1})_\mu ^\nu P_\nu.
\end{eqnarray}
These equations are simply one way of writing the Poincar\'{e} commutation
relations in which the relations of the four-momentum operators are
emphasized.  The mass operator is
defined to be
$M:=\sqrt{P\cdot P}$ and must have a nonnegative spectrum.

Since the four-momentum operators are the generators of space-time
translations, they can be used to define the relativistic
generalization of the time dependent Schr\"{o}dinger equation,
\begin{eqnarray}
i\frac{\partial \Psi _x}{\partial x^\mu}&=&P_\mu \Psi_x
\end{eqnarray}
where $\Psi_x$ is an element of the Hilbert space and $x (=x_\mu)$ is a
space-time point.  From Eq.3 it follows that $\Psi_x$ satisfies a
generalized Klein-Gordan equation,
\begin{eqnarray}
(\frac{\partial}{\partial x^\mu }\frac{\partial}{\partial x_\mu}+ M^2)
\Psi_x&=&0,
\end{eqnarray} 
where $M$ is the mass operator.  It is Eqs. 3,4 that take the place of
equations of motion. If $P_\mu$ has no explicit space-time dependence,
then Eq.3 can be written as an eigenvalue equation for the four-momentum
operator: 
\begin{eqnarray}
P_\mu \Psi&=&p_\mu \Psi.
\end{eqnarray}

Assume now that $P_\mu$ is the sum of hadron, photon, and electromagnetic
four-momentum operators, and satisfies Eqs.1,2.  The hadron four-momentum
operator, $P_\mu(h)$, is assumed to be independent of photon
creation and annihilation operators, while the photon four-momentum
operator, $P_\mu(\gamma)$, is proportional to $c^\dagger c$ (see
Eq.23, reference \cite{e}).  The main point of this paper is to show how
to construct a conserved current operator for the electromagnetic vertex in
the presence of hadronic interactions. 

 In particular, if the hadronic four-momentum operator is of
Bakamjian-Thomas form, which in the point form means that the four-momentum
operator is the product of a mass and four-velocity operator, then, as will
be shown in section 5, matrix elements of the model dependent current can
be written as a commutator of the interacting mass operator with a freely
chosen current operator.

\section{Point Form Electrodynamics}
The fundamental quantities in point form electrodynamics needed to
construct many-body currents are the four-momentum operator, $P_\mu$, and
the four-vector potential operator at the space-time point zero, $A_\mu(0)$
(which is always kinematic and defined via the photon creation and
annihilation operators, see Eq.26, reference
\cite{e}).  From these quantities the electromagnetic field tensor at the
space-time point zero is defined as
\begin{eqnarray}
iF_{\mu \nu}(0):=[A_\mu(0),P_\nu]-[A_\nu(0),P_\mu],
\end{eqnarray}
and is antisymmetric in $\mu$ and $\nu$.  From this definition it follows
that the electromagnetic field tensor at the space-time point $x$
satisfies the usual relationship with the four-vector potential:
\begin{eqnarray}
iF_{\mu \nu}(x):&=&ie^{iP\cdot x}F_{\mu \nu }(0)e^{-iP\cdot x}\\
&=&[A_\mu(x),P_\nu]-[A_\nu(x),P_\mu]\\
&=&i(\frac{\partial A_\mu(x)}{\partial x^\nu} -\frac{\partial A_\nu
}{\partial x^\mu}),
\end{eqnarray}
where $A_\mu(x)=e^{iP\cdot x}A_\mu(0)e^{-iP\cdot x}$.  $A_\mu(0)$ is a
linear combination of photon creation and annihilation operators and hence
acts only on the photon Fock space.  The correct Lorentz transformation
properties of all the field quantities follow from the Lorentz
properties of operators at the space-time point zero.

A electromagnetic current operator at the space-time point zero can then
be defined as
\begin{eqnarray}
iJ_\mu(0):&=&[F_{\mu \nu}(0),P^\nu],
\end{eqnarray}
from which it follows that
\begin{eqnarray}
iJ_\mu(x)&=&ie^{iP\cdot x} J_\mu(0) e^{-iP\cdot x}\\
&=&e^{iP\cdot x}[F_{\mu \nu}(0),P^\nu]e^{-iP\cdot x}\nonumber\\
&=&i\frac{\partial F_{\mu \nu}(x)}{\partial x_\nu}.
\end{eqnarray}
Such a current is always conserved.  This follows directly from the
antisymmetry of $F_{\mu\nu}(x)$ and Eq.12, but it is worthwhile
deriving current conservation as a commutator relation, namely 
$[P^\mu,J_\mu(0)]=0$, since this form of current conservation will be
used in the following section.  
 
Now,
$\partial J_\mu(x)/\partial x_\mu=0$ is equivalent to
$[P^\mu,J_\mu(0)]=0$.  But
\begin{eqnarray}
i[P^\mu,J_\mu(0)]&=&[P^\mu,[F_{\mu \nu}(0),P^\nu]]\nonumber\\
&=&-[F_{\mu\nu}(0),[P^\nu,P^\mu]]-[P^\nu,[P^\mu,F_{\mu \nu}(0)]]\nonumber\\
&=&[P^\nu,[F_{\mu\nu}(0),P^\mu]]\nonumber\\&=&-[P^\mu,[F_{\mu\nu}(0),P^\nu]]\nonumber\\
&=&0.
\end{eqnarray}
To arrive at this result the Jacobi identity for three operators was
used, as well as the antisymmetry of $F_{\mu\nu}(0)$ and the point form
Eq.1.  Thus, any current operator defined by Eq.10 is conserved if the
electromagnetic field tensor is antisymmetric and the four-momentum
operators commute with one another.

Eq.12 turns around the usual relationship between field and current.  In
classical electrodynamics, Eq.12 is usually interpreted as a differential
equation for the field in the presence of specified currents.  In this
paper Eq.10 will define a many-body current, given  $F$ and the
four-momentum operator.  Moreover, such a current is not only conserved,
but also carries zero charge.  That is, matrix elements of the charge
operator on eigenstates of the four-momentum operator are zero: 
\begin{eqnarray}
i<p^{'} ...|\hat{Q}|p...>&=&i\int d^3 x<p^{'}...|J_{\mu=0}(\vec{x})|p...>\\
&=&i\int d^3 x<p^{'}...|e^{i\vec{P}\cdot \vec{x}}
J_{\mu=0}(0)e^{-i\vec{P}\cdot\vec{x}}|p...>\nonumber\\
&=&\int d^3 x
e^{i(\vec{p}^{'}-\vec{p})\cdot\vec{x}}<p^{'}...|[F_{0\nu}(0),P^{\nu}]|p...>
\nonumber\\
&=&\delta^3(\vec{p}^{'}-\vec{p})(p-p^{'})^{\nu}<p^{'}...|F_{0\nu}(0)|p...>
\nonumber\\
&=&0;
\end{eqnarray} 
 here $|p...>$ is an eigenstate of the four-momentum operator
(the dots indicate other quantum numbers not needed in the derivation) and
it is assumed that the two eigenstates have the same mass.

Because fields, in general, are not local, gauge transformations are
defined by adding an operator (rather than c-number) to $A_\mu(x)$:
\begin{eqnarray}
A_\mu^{'}(0):&=&A_\mu(0)+i[P_\mu,\chi(0)],
\end{eqnarray}
where $\chi(0)$ is a Lorentz scalar operator.  If $\chi(0)$ is space-time
translated to $\chi(x)$, it follows that
\begin{eqnarray}
A_\mu^{'}(x)&=&A_\mu(x)+\frac{\partial \chi(x)}{\partial x^\mu},\\
iF_{\mu\nu}^{'}(0):&=&[A_{\mu}^{'}(0),P_\nu]-[A_{\nu}^{'}(0),P_\mu]\\
&=&iF_{\mu\nu}(0),\nonumber
\end{eqnarray}
so that the electromagnetic field tensor is gauge invariant.  Since the
gauge transformation defined in Eq.16 involves a scalar operator rather
than a c-number function, as is the case for local fields, the link to
gauge transformation in terms of photon creation and annihilation
operators is different than that given in reference \cite{e}.  The link
is given by writing $\chi(0)$ in terms of momentum dependent operators:
\begin{eqnarray}
\chi(0)&=&\int \frac{d^3 k}{k_0} (\tilde{c}(k)+\tilde{c}(k)^\dagger),\
\end{eqnarray}
where $\tilde{c}(k)^\dagger$ is the adjoint of the operator $\tilde{c}(k)$
and guarantees that $\chi(0)$ is hermitian.  Using the results of reference
\cite{e}, Eq.31, where $A_\mu(0)$ ($A_\mu^{'}(0)$) is expanded in terms of
photon creation and annihilation operators, and keeping only the
annihilation operator parts, Eq.16 becomes
\begin{eqnarray}
-\epsilon_\mu(k,\alpha)g_{\alpha,\alpha}(c^{'}(k,\alpha)-c(k,\alpha))&=&
i[P_\mu,\tilde{c}(k)]\nonumber\\
&=&-k_\mu f(k)I.\nonumber\\
{[P_\mu,\tilde{c}(k)]}&=&ik_\mu f(k)I
\end{eqnarray}
Since the four-momentum operator $P_\mu$ contains all the dynamics, Eq.18
implies that the gauge operator 
$\tilde{a}(k)$ also is dynamical, while the c-number gauge
transformation
$f(k)$ given in Eq.31 of reference
\cite{e} remains kinematical.  $\epsilon_\mu(k,\alpha)$ is the polarization
vector.

To conclude this section Maxwell's equations are shown to be expectation
values of the operator relations.  Thus,
\begin{eqnarray}
<F_{\mu\nu}(x)>&=&(\Psi_0,e^{iP\cdot x} F_{\mu\nu}(0)
e^{-iP\cdot x}\Psi_0)\nonumber\\
&=&(\Psi_x ,F_{\mu\nu}(0) \Psi_x),\\
\frac{\partial}{\partial x_{\nu}}<F_{\mu\nu}(x)>
&=&(\Psi_x,J_\mu(0)\Psi_x)\nonumber\\
&=&<J_\mu(x)>;\\
<F_{\mu\nu}(x)>&=&\frac{\partial}{\partial
x^\nu}<A_\mu(x)>-\frac{\partial}{\partial x^\mu}<A_\nu(x)>,
\end{eqnarray}
where use has been made of the relativistic Schr\"{o}dinger equation, Eq.3
and
$\Psi_0$ is a wavefunction at the space-time point zero. Now in
general\\
$\frac{\partial}{\partial x_\mu} <A_\mu (x)>\neq0$.  But, just as in
classical electrodynamics, where one may make a gauge transformation so
that the four divergence of the vector potential is zero, so too the
operator
$\chi(0)$ can be chosen so that the delambertian of expectation values of
$\chi (x)$ produces a four-divergence of expectation values of the
four-vector potential being zero.  The point is that for any four-momentum
operator satisfying the point form equations, Eqs.1,2, an operator gauge
transformation, Eq.16 results in a gauge invariant field operator,
$F_{\mu \nu}(0)$, as seen in Eq.18, and a gauge invariant current
operator.
\section{Dynamically Determined Many-Body Currents}
The goal of this paper is to construct  conserved electromagnetic current
operators on the hadronic Hilbert space as the sum of one-body
current operators defined in the previous paper \cite{a}, and a many-body
current operator,  
\begin{eqnarray}
J_{\mu}(0):&=&J_{\mu}^{1}(0)+J_{\mu}^{many-body}(0),
\end{eqnarray}
itself the sum of dynamically determined and model dependent pieces.  
In this section the dynamically determined currents will be built from the
free current operators constructed in reference \cite{a}.  As previously
noted, such operators are conserved for the free hadronic four-momentum
operator, but not in the presence of hadronic interactions.  This can be
seen most clearly by writing translational covariance as
\begin{eqnarray}
J_\mu(x)&=&e^{iP\cdot x} J_\mu(0) e^{-iP\cdot x}\\
{[P_\mu ,J_\nu(x)]}&=&-i\frac{\partial J_\nu(x)}{\partial x^\mu};
\end{eqnarray}
$J_\nu(x)$ is clearly conserved if $[P^\mu ,J_\mu(0)]=0$.  This is the case
if the four-momentum operator is the free four-momentum operator and the
current is a one-body current.  However, if the (hadronic) four-momentum
operator is the sum of free and interacting four-momentum operators,
 $P^\mu(h) =P^\mu(fr)+P^\mu(I)$,
(satisfying Eqs.1,2), current conservation will not hold unless a term
(denoted by "DD" for dynamically determined) is added to the one-body
current operator:
\begin{eqnarray}
{[P^\mu(fr)+P^\mu(I),J_\mu ^1 (0)+J_\mu^{DD}(0)]}&=&0.
\end{eqnarray}
If the interacting four-momentum operator, $P^\mu(I)$ does not commute with
the one-body current operator, $J_\mu^1(0)$, there must be additional
(dynamically determined) currents present in order to satisfy current
conservation, exactly as in the non-relativistic case.

 Making use of the fact that the one-body current operator is
conserved with respect to the free four-momentum operator,
$[P^\mu(fr),J_\mu ^1(0)]=0$, and taking eigenfunctions of the hadronic
four-momentum operator to construct current operator matrix elements, Eq.27
becomes
\begin{eqnarray}
Q^\mu<p^{'}|J_\mu^{DD}(0)|p>&=&-<p^{'}|[P^\mu(I),J_\mu^1(0)]|p>,
\end{eqnarray}
where $Q^\mu$ is the four-vector momentum transfer, $Q^\mu=(p^{'}-p)^\mu$.

A general solution to Eq.28 is given by
\begin{eqnarray}
<p^{'}|J_\mu
^{DD}(0)|p>&=&-\frac{(Q_\mu+Q_\mu^{\perp})}{Q^2}
<p^{'}|[P^\nu(I),J_\nu^1(0)]|p>\\
&=&-\frac{(Q_\mu+Q_\mu^{\perp})}{Q^2}Q^\nu<p^{'}|J_\nu^1(0)|p>;
\end{eqnarray}
here $Q^{\perp}$ is a four-vector perpendicular to $Q, Q\cdot
Q^{\perp}=Q^\mu Q_\mu^{\perp}=0$, and is determined by the requirement that
there should be no pole at $Q^2=0$.  As can be seen from Eq.29, if the
interacting four-momentum operator is zero, the dynamically determined
current operator also is zero.

To see the effects of adding a dynamically determined many-body current
operator to the free current operator to ensure current conservation, it is
useful to go to a (generalized) Breit frame, where 
$p^{'}(st)=m^{'}(ch\alpha,0,0,sh\alpha),\\ p(st)=m(ch\alpha,0,0,-sh\alpha)$,
and $Q(st)=((m^{'}-m)ch\alpha,0,0,(m^{'}+m)sh\alpha)$.  The appendix shows
that if $Q^{\perp}=-((m^{'}+m)sh\alpha,0,0,(m^{'}-m)ch\alpha)$ there will be
no pole at
$Q^2=0$. Note that if the masses of the initial and final state are the same
($m^{'}=m$), the four-vector momentum transfer is the usual one, with only
the third component nonzero. Then the third component of the total current
in the Breit frame is zero, because the contribution from the one-body
current matrix element is exactly balanced by the dynamically determined
current, as seen in Eq.30.

More generally the current operator is
\begin{eqnarray}
J_\mu(0)&=&J_\mu^1(0)+J_\mu^{DD}(0);\\
<p^{'}|J_\mu(0)|p>&=&<p^{'}|J_\mu^1(0)|p>-\frac{Q_\mu+Q_\mu^{\perp}}{Q^2}
Q^\nu<p^{'}|J_\nu^1(0)|p>,
\end{eqnarray}
and from Eq.32 it follows that $J_\mu(0)$ is conserved,
 $Q^\mu <p^{'}|J_\mu(0)|p>=0.$

Moreover the form of $J_\mu^{DD}(0)$ in Eq.30 implies it carries
charge. Using Eq.14 for the charge operator, the relevant matrix element to
be evaluated is $<p|J_{\mu=0}^{DD}(0)|p>$;  since such matrix elements in
the point form are Lorentz covariant, it is most convenient to use the Breit
frame and set  $\alpha=0$.  Then
\begin{eqnarray}
<p^{'}(st)|J_{\mu=o}^{DD}(0)|p(st)>|_{\alpha
=0}&=&-\frac{(Q(st)+Q^{\perp}(st))_{\mu=0}}
{Q^2}\nonumber\\&&Q^\nu<p^{'}(st)|J_\nu^1(0)|p(st)>|_{\alpha =0}\\
&=&-<p^{'}(st)|J_{\nu=3}^1(0)|p(st)>|_{\alpha=0},
\end{eqnarray}
so that the total charge content is given by
\begin{eqnarray}
<p^{'}(st)|J_{\mu=0}(0)|p(st)>&=&<p^{'}(st)|J_{\mu=0}^1(0)
-J_{\mu=3}^1(0)|p(st)>
\end{eqnarray}
for $\alpha=0.$

To see the implications for the total charge as a sum of constituent charges
consider the simple model in which a spinless bound state of mass m is made
from n spinless constituents of mass $m_c$.  The goal is to compute the
one-body matrix elements in Eq.35 for $\mu=0$ and $\mu=3$, by assuming that
the current matrix element of the first constituent has the form
$<p_1^{'}|J_\mu(0)|p_1>=(p_1^{'}+p_1)_\mu f_1((p_1^{'}-p_1)^2)$, with
$f_1()$ an arbitrary constituent form factor, whose value at 0 is the
charge of the first constituent.  The other constituents will have similar
current matrix elements.

To carry out such a computation it is convenient to switch from single
particle constituent variables, $p_i, i=1...n$ to internal variables, $k_i$,
with $\sum \vec{k_i}=\vec{0}$; the connection is given by $p_i=B(v)k_i$,
where $B(v)$ is a  boost Lorentz transformation taking the four-vector
$(1,0,0,0)$ to the four-velocity $v$ ($v\cdot v=v^\mu v_\mu=1)$.  The
Jacobian
$\mathcal{J}$which converts the relativistic measure $\prod \frac{d^3
p_i}{2E_i}$ to a measure in velocity and internal momentum variables is
given by
\begin{eqnarray}
\prod \frac{d^3 p_i}{2E_i}&=&\frac{d^3 v}{v_0}\mathcal{J}\prod^{'} d^3 k_i\\
\mathcal{J}&=&\frac{M_0}{\prod 2\omega_i},
\end{eqnarray}
where the prime on the product in Eq.36 indicates a product over $n-1$
variables rather than $n$ variables, $M_0$ is the total n-particle mass,
$M_0=\sqrt{\sum p_i\cdot \sum p_i}=\sum \omega _i$, and $\omega_i$ is the
energy of the 
$i^{th}$ constituent, $\omega_i=\sqrt{m_c^2+\vec{k_i}^2}$.  If
$\Psi(\vec{k_i})$ is the bound state wavefunction for the particle of mass
$m$, normalized to $||\Psi||^2=\int \mathcal{J}\prod^{'} d^3 k_i
|\Psi(\vec{k_i})|^2=1$, then the desired current matrix element can be
written as
\begin{eqnarray}
<p^{'}(st)|J_\mu^1 (0)|p(st)>&=&\int \mathcal{J^{'}} \prod^{'} d^3 k_i^{'}
\mathcal{J}\prod^{'}d^3 k_i \Psi^{\ast}(k^{'}_i)\nonumber\\
&&<v^{'}(st)k_i^{'}|J_\mu^1(0)|v(st)k_i>\Psi(k_i);\\
<v^{'}(st)k_i|J_\mu^1(0)|v(st)k_i>&=&
\prod^{'}2\omega_i\delta^3(k_i^{'}-B^{-1}(v^{'}(st))B(v(st))k_i)\nonumber\\
&&(p^{'}_1+p_1)_\mu f_1((p_1^{'}-p_1)^2)\mathcal{F}.
\end{eqnarray}
Eq.39 is the matrix element when only constituent 1 has been struck; to
get the total charge similar matrix elements for all the other
constituents must be added to Eq.39.   The factor
$\mathcal{F}$ is needed to ensure that for $\alpha=0$ the current matrix
element, Eq.38 gives the correct total charge.  That is, when $\alpha=0$,
$v^{'}(st)=v(st)=(1,0,0,0)$ and the current matrix element for constituent
1 being struck is given by
\begin{eqnarray}
<v^{'}(st)k_i|J_\mu^1(0)|v(st)k_i>|_{\alpha=0}&=&f_1(0)\int\mathcal{J}
\prod^{'}d^3k_i|\Psi(k_i)|^2\nonumber\\
&&\frac{M_0}{2\omega_1}2(k_1)_\mu\mathcal{F}.
\end{eqnarray}

The requirement on the charge operator, Eq.35, is that the charge of the
bound state particle should equal the sum of the constituent charges.  If
there were no dynamically determined currents, Eq. 35 would say that only
the zero component of the one-body current matrix element contribute;  in
that case, $\mathcal{F}=1/M_0$.  However, the current in Eq.35 is the sum of
one-body and dynamically determined currents, which means the factor must
include both the zero and third component;  in that case  
$\mathcal{F}=\frac{\omega_1}{M_0(\omega_1-k_{1z})}$. 
Thus it is clear that the dynamically determined current carries charge and
must be taken into account in order that the total charge be the sum of the
constituent charges.

\section{Model Dependent Currents}
The modification of $J_{\mu}^1$ to ${J}_\mu=J_\mu^1+J_\mu^{DD}$  corresponds
to the addition of a dynamically determined term to the free current
operator, analogous to the model independent current in non-relativistic
quantum mechanics.  The goal of this section is to construct the analogue
of model dependent currents in nonrelativistic quantum mechanics, by using
the current operators defined in section 3.  Such currents, it will be
recalled, are defined as the commutator of an antisymmetric operator
with the hadronic four-momentum operator (see Eq.10); though the
antisymmetric operator in Eq.10 is the electromagnetic field tensor, the
only property used in showing that the current operator is conserved is
that the operator in the commutator be antisymmetric.  Moreover Eq.14 shows
that the current also carries no charge.  These results are the point form
analogue of the classical electrodynamics result,
namely that it is always possible to add to the current a term of the form
$\frac{\partial A^{\mu\nu}(x)}{\partial x^\nu}$ (where $A^{\mu\nu}(x)$ is an
antisymmetric field) that is conserved and carries no charge
\cite{f}.

When $F_{\mu\nu}$ is replaced by an arbitrary antisymmetric operator, it
should satisfy a cluster requirement, that when
the hadronic interactions are zero, $A_{\mu\nu}$ should also be zero.  That
is, there should be no model dependent currents when the interacting
hadronic mass operator is zero.  Such a requirement is readily satisfied by
writing the field tensor as
\begin{eqnarray}
iA_{\mu\nu}:&=&[I_\mu,P_\nu(I)]-[I_\nu,P_\mu(I)],
\end{eqnarray}
where $P_\mu(I)$ is the interacting hadronic four-momentum
operator and
$I_{\mu}$ is any (current) operator that transforms as a
four-vector.  It need not be conserved and can carry charge; the model
dependent current operator that it generates will of course be conserved and
carry no charge.  Examples of $I_{\mu}$ are the original one-body
current operator, with or without additional form factors, as well as any
Feynman diagram with an external photon line, viewed as the kernel of a
current operator.

Matrix elements of the model dependent (MD) current operator are given by
\begin{eqnarray}
<p^{'}...|J_\mu^{MD}|p...>&=&Q^{\nu}<p^{'}...|[I_\mu,V_\nu
M_I]|p...>\nonumber\\
&&-Q^{\nu}<p^{'}...|[I_\nu,V_\mu M_I]|p...>,
\end{eqnarray}
which can be simplified by going to the Breit frame:
\begin{eqnarray}
<p^{'}(st)...|J_\mu^{MD}|p(st)...>&=&-2msh^2\alpha<p^{'}(st)...
|[I_\mu,M_I]_+|p(st)...>\nonumber\\
&&+2msh\alpha ch\alpha\nonumber\\
&&<p^{'}(st)...|[M_I,I_3]|p(st)...>\delta_{\mu 0},
\end{eqnarray}
for $\mu=0,1,2$;  for $\mu=3$ the matrix element is zero.  To see the
effects of model dependent currents on form factors requires choosing a
current $I_\mu$ and then calculating integrals of the form given in Eq.38;
this will not be done here.

\section{Conclusion}
In this paper a general procedure for constructing conserved current
operators has been given.  The total current operator is written as the sum
of three operators, a free current operator plus two many-body
operators.  The free current operator must be modified because it is not
conserved with respect to the hadronic interactions.  As shown in section
4 the procedure for modifying it is dynamically determined in the sense that
no new operators are invoked to make it conserved. 
The current operator of section 5 is, on the other hand, highly model
dependent, in that many different operators can be chosen to generate the
field tensor in Eq.41, which then generates the model dependent conserved
current.

Though  the dynamically determined current
follows from current conservation, as is also the case in nonrelativistic
quantum mechanics, the form it takes (see Eq.30 ) has been used in the
past to attain current conservation phenomenologically \cite {g}.  What has
been shown in section 4 is that such a correction is incomplete, in that
another term, the
$Q^{\perp}_\mu$ in Eq.30  is also needed, to guarantee that there are no
poles in the dynamically determined current.  This in turn leads to the
possibility that the dynamically determined current carries charge, which
was investigated by looking at a simple bound state model.  This model shows
that indeed the dynamically determined current carries charge, and must be
taken into account, in order that the total charge equal the sum of the
constituent charges. 

The motivation for this work comes from calculations done for elastic
deuteron \cite{h} and nucleon form factors \cite{i}, which give very
different results when compared to data.  In both calculations the point
form of relativistic quantum mechanics is used to compute matrix elements
of free current operators in the Breit frame \cite{c}.  In the deuteron
calculation the A structure function is below the data already at momentum
transfers of 2 Gev$^2$.  Clearly many-body currents are needed to get better
agreement with data.  On the other hand, using free current operators for
constituent quarks with no anomalous magnetic moments or other internal
structure,  Breit frame matrix elements give very good agreement with data,
except for some static properties of the nucleons.

In order to try to understand these results it is necessary to construct
many-body currents that change the deuteron form factors, but not the
nucleon form factors.  These many-body currents should be conserved, but
not renormalize the charge.  It is exactly such many-body currents that
have been constructed in this paper.  Since the interacting hadronic mass
operator has been chosen in such a way as to give  wave
functions whose eigenvalues are in good agreement with data, and the
dynamically determined currents are fixed by the one-body currents, the only
way in which different model currents can be introduced to see how they
influence the deuteron and nucleon wave functions is through the model
dependent currents, given in Eq.41 by the current $I_\mu$.  Actual
calculations of dynamically determined and model dependent currents for
nucleon and deuteron form factors remain to be carried out.
\section{appendix}
In this appendix matrix elements of the dynamically determined current
operator are analyzed to see if there is a pole at $Q^2=0$ when the
initial and final four-momenta may have different masses (inelastic form
factors).  In such a situation there is no longer a unique definition of a
Breit frame, but the generalization chosen here reduces to the usual
definition when the initial and final masses are the same.  From Eq.30 the
matrix elements of the dynamically determined currents are

\begin{eqnarray}
<p^{'}...|{J}_\mu^{DD}|p...>&=&-\frac{(Q_\mu+Q_\mu^{\perp})Q^{\nu}}{Q^2}
<p^{'}...|J_\nu^{1}|p...>,
\end{eqnarray}
with $Q=p^{'}-p$.  If $Q\cdot Q^{\perp}=0$, then the sum of one-body plus
dynamically determined current operators is conserved.  Since the
initial and final masses are no longer equal, the question is whether
$Q^{\perp}$ can be chosen in such a way to ensure that there is no pole at
zero momentum transfer \cite{g}.

To see how this is possible, choose a generalized Breit frame with
$p^{'}(st)=m^{'}(ch\alpha,0,0,sh\alpha)$ and
$p(st)=m(ch\alpha,0,0,-sh\alpha)$;  this reduces to the usual
definition of a Breit frame when the two masses are equal.  With these
standard momenta, the momentum transfer squared is
$Q^2(st)=(m^{'}-m)^2-4m^{'}msh^2\alpha$ and is zero when
$sh\alpha=\frac{m^{'}-m}{2\sqrt{m^{'}m}}$, which reduces to the elastic
case when the masses are the same.

The goal now is to choose $Q^{\perp}(st)$ so that $Q(st)\cdot
Q^{\perp}(st)=0$ and there is no pole at $Q^2(st) =0$ \cite{h}.  The
condition that
$Q(st)\cdot Q^{\perp}(st)=0$ means that
\begin{eqnarray}
(Q+Q^{\perp})(st)&=&((m^{'}-m)ch\alpha,0,0,(m^{'}+m)sh\alpha)\nonumber\\
&&+a((m^{'}+m)sh\alpha,0,0,(m^{'}-m)ch\alpha)
\end{eqnarray}
where $a$ is a constant to be chosen so that $(Q+Q^{\perp})(st)=0$ at $Q^2
(st)=0$.  Write $\alpha=\alpha_0$ at $Q^2(st)=0$.  Then
$sh\alpha_0=\frac{m^{'}-m} {2\sqrt{m^{'}m}}$,
$ch\alpha_0=\frac{m^{'}+m}{2\sqrt{m^{'}m}}$, and if $a=-1$,
$Q+Q^{\perp}(st)=0$ at $\alpha =\alpha_0$;  thus the numerator and
denominator are both zero at
$Q^2(st)=0$ ($\alpha=\alpha_0$).  Moreover, if the numerator and denominator
are both expanded about $\alpha=\alpha_0$, it is readily checked that the
ratio is finite for $m^{'}\neq m$.  The final result is that if
\begin{eqnarray}
(Q+Q^{\perp})(st)&=&[(m^{'}-m)ch\alpha-(m^{'}+m)sh\alpha](1,0,0,-1)
\end{eqnarray}
the modified current in Eq.30 is conserved and has no pole at $Q^2=0$.

\end{document}